%% file: main.tex
\def\year{2021}\relax
\documentclass[letterpaper]{article} 
\usepackage{aaai21}  
\usepackage{times}  
\usepackage{helvet} 
\usepackage{courier}  
\usepackage[hyphens]{url}  
\usepackage{graphicx} 
\usepackage{natbib}  
\usepackage{caption} 
\usepackage{amsmath,amssymb,amsthm,amsfonts}
\usepackage{algorithm, algorithmic}
\usepackage{mathtools}

\title{S++: A Fast and Deployable Secure-Computation Framework for Privacy-Preserving Neural Network Training} 
\author {
    Prashanthi Ramachandran \textsuperscript{\rm 1}
    Shivam Agarwal \textsuperscript{\rm 1}
    Arup Mondal \textsuperscript{\rm 1} \\
    Aastha Shah \textsuperscript{\rm 1} 
    Debayan Gupta \textsuperscript{\rm 1} \\
}
\affiliations {
    \textsuperscript{\rm 1} Ashoka University \\
    prashanthi.r@alumni.ashoka.edu.in, shivam.agarwal@alumni.ashoka.edu.in,
    arup.mondal\_phd19@ashoka.edu.in,\\ aastha.shah@alumni.ashoka.edu.in,
    debayan.gupta@ashoka.edu.in 
}

\begin{document}

\maketitle

\begin{abstract}
We introduce S++, a simple, robust, and deployable framework for training a neural network (NN) using private data from multiple sources, using secret-shared secure function evaluation. In short, consider a virtual third party to whom every data-holder sends their inputs, and which computes the neural network: in our case, this virtual third party is actually a set of servers which individually learn nothing, even with a malicious (but non-colluding) adversary. 

Previous work in this area has been limited to just one specific activation function: ReLU, rendering the approach impractical for many use-cases. For the first time, we provide fast and verifiable protocols for \textbf{all} common activation functions and optimize them for running in a  secret-shared manner. The ability to quickly, verifiably, and robustly compute exponentiation, softmax, sigmoid, etc., allows us to use previously written NNs without modification, vastly reducing developer effort and complexity of code. In recent times, ReLU has been found to converge much faster and be more computationally efficient as compared to non-linear functions like sigmoid or tanh. However, we argue that it would be remiss not to extend the mechanism to non-linear functions such as the logistic sigmoid, tanh, and softmax that are fundamental due to their ability to express outputs as probabilities and their universal approximation property. Their contribution in RNNs and a few recent advancements also makes them more relevant.

\end{abstract}


\input{introduction}
\input{background}
\input{protocol}
\input{evaluation}
\input{conclusion}


\begin{quote}
\begin{small}
\bibliography{references.bib}
\end{small}
\end{quote}


\input{appendix}

\end{document}

%% file: introduction.tex
\section{Introduction}\label{section:intro}


Neural networks (NN) are used in areas ranging from image classification to machine translation, and there are often multiple parties that contribute possibly private data to the training process. However, training data interaction and the resulting model may still leak a significant amount of private data. Thus, there arises a need for \textit{secure} neural networks, which can help parties collaboratively train a neural network without revealing their private input data. 

Several approaches have been proposed to overcome possible data leakage, based on secure multi-party computation (MPC). MPC is a powerful way to protect sensitive training data which uses a range of cryptographic primitives to allow multiple parties to compute a function without revealing the inputs of any individual party (beyond what is implied by the output of the function itself). However, previously proposed MPC-based schemes~\cite{ohrimenko2016oblivious,hunt2018chiron,mohassel2017secureml,wagh2018securenn,patra2020blaze} do not support models that use exponentiation-based activation functions such as \textit{logistic sigmoid}, \textit{softmax}, and \textit{tanh}. These protocols are largely restricted to ReLU and its variants, which might restrict applicability in some specific cases (as described in the motivation section). 

In this paper, we propose \textit{S++}, a three-party secure computation framework for secure exponentiation, and exponentiation-based activation functions such as logistic sigmoid, tanh, and softmax. This enables us to construct three-party secure protocols for training and inference of several NN architectures such that no single party learns any information about the data. \textit{S++} is an efficient MPC-based privacy-preserving neural network training framework based on ~\cite{wagh2018securenn} for 3PC with the activation functions \textit{logistic sigmoid}, \textit{tanh}, their derivatives, and \textit{softmax}. In our setting, there are $D$ (where $D$ can be arbitrary) data owners who wish to jointly train a model over their data with the help of $3$-servers. In the setup phase, these data owners create "additive secret-shares" of their input data and send one share each to $2$-servers. In the computation phase, the $3$-servers (the $2$ primary servers and the helper server) collectively run an interactive protocol to train a neural network on the data owners' data without learning any information beyond the trained model.


\subsection{Contributions}\label{subsection:contribution}
We summarize our key contributions as follows:
\begin{itemize}

    \item We first propose a secure protocol for exponentiation in the three-party setting. This protocol is based on SCALE MAMBA's~\cite{aly2020scale} protocol for base 2 exponentiation. It also uses primitives from \cite{catrinasaxena}. We modify these ideas for our three-party setting and additively secret-shared data.
        
    \item We describe novel secure protocols for the \textit{logistic sigmoid}, \textit{softmax}, and \textit{tanh}---popular activation functions that are significantly more complex than ReLU (described by~\cite{wagh2018securenn})---along with their derivatives with the help of the above-mentioned exponentiation protocol. The inclusion of these protocols in a secure and private setting vastly increases the practicality of the framework and enables people to convert their protocols into secure ones without having to redesign the actual internal structure of their NNs.
    
\end{itemize}

\subsection{Organization of the paper}
The remainder of this paper is organized as follows: we first go over our \textit{motivation} for extending secure protocols to functions such as the logistic sigmoid, tanh and softmax. Then, we look at \textit{recent work} in this area, largely in the realm of MPC. After that, we describe the \textit{notations} used in our protocols and explain some \textit{cryptographic primitives}. In the \textit{Protocols} section, we delve into our \textit{architecture} and the \textit{supporting protocols}, specifically the supporting 3-server protocols that serve as building blocks for our main protocols. We describe our \textit{main protocols}: exponentiation, logistic sigmoid, tanh, their derivatives, and softmax. After that, we describe the theoretical efficiency of our protocols and provide an \textit{evaluation} of our experiments. Lastly, we describe the future plans and directions for this work.

\subsection{Motivation}\label{subsection:motivation}

\subsubsection{Logistic sigmoid and variants}

The logistic sigmoid is a common activation function to introduce non-linearity, where sigmoid$(x) = \displaystyle\frac{1}{1 + e^{-x}}$. It essentially squashes a real value in [0, 1] and has been used widely over the years for its simplicity, especially in binary classification tasks with a maximum likelihood loss function. It also has important variants like the tanh and softmax:

\begin{itemize}
    \item \textbf{Tanh as a rescaled logistic sigmoid: }The tanh is a rescaling and stretching of the logistic sigmoid that maps to [-1, 1] i.e., tanh$(x)$ = 2$\times$sigmoid$(2x) - 1$. Extending the protocol to the tanh facilitates applications in recurrent neural networks where it is widely used.

\item \textbf{Logistic sigmoid as a special case of the softmax: }The softmax is often used for classification tasks, where softmax$(z_i) = \displaystyle\frac{e^{z_i}}{\sum_{i=1}^k e^{z_i}}$. The logistic sigmoid happens to be a special case of this function, often used for binary classification. Enabling softmax thus also extends the framework to multi-class classification problems.
\end{itemize}

\subsubsection{Why is ReLU not enough?}
\label{subsubsection:relu}

Despite ReLU being more computationally efficient than logistic sigmoid or tanh \cite{krizhevsky2017imagenet}, we argue that it is worth extending the secure computation protocol to popular functions \textit{particularly} involving exponentiation such as logistic sigmoid, tanh, and softmax for the following reasons (with more details in the appendix):
\begin{itemize}
    \item For classification tasks, output layers usually use softmax or logistic sigmoid because they make for more interpretable values in [0, 1] as opposed to ReLU and variations that are unbounded. In fact, though previously missing in the literature, \cite{asadi2020approximation} provide theoretical proofs on the universal approximation of softmax used in the output layer for multiclass classification tasks as well.
    \item ReLU's unbounded nature limits its applicability in RNNs. Capturing long-term dependencies becomes harder with an unbounded function due to exploding gradients, as explored in \cite{pascanu2013difficulty}. Because of the temporal correlations captured by RNNs (by forming connections between hidden layers), long-term components add up and the norm of the gradient of the hidden layers \textit{explode} and ReLU's unbounded nature exacerbates this problem. There have been a few notable improvements to assuage this, but even in LSTMs, the tanh often gives more consistent results \cite{kent2019performance} and ReLU tends to require greater fine-tuning.
    \item 'Leaky' ReLU has been used to overcome ReLU's vanishing gradient problem. \cite{xu2016revise} have also revised logistic sigmoid to a scaled version of it and tanh to a `leaky' tanh. The `leaky' version penalizes the negative part of the domain. They found it to be much faster than the tanh, it gave better results than ReLU, and is almost identical to leaky ReLU. Such potential improvements highlight the need for more secure protocols for such functions.
\end{itemize}
Though ReLU has become widely adapted in the past few years, there are still some tasks wherein sigmoids are preferred over ReLU, and it is worthwhile to extend such secure protocols for these as well. 

\subsection{Recent Work}\label{subsection:relatedwork}

There have been many privacy-preserving machine learning (PPML) frameworks for various situations, such as Decision Trees~\cite{agrawal2000privacy,lindell2000privacy}, Linear Regression~\cite{du2001privacy,sanil2004privacy}, k-means clustering~\cite{jagannathan2005privacy,bunn2007secure}, SVM classification~\cite{yu2006privacy,vaidya2008privacy}, and logistic regression~\cite{slavkovic2007secure}. However, these cannot generalize to a number of (complex and high accuracy) standard machine learning applications.

To overcome this, SecureML~\cite{mohassel2017secureml} provided secure protocols for linear regression, logistic regression, and neural networks with linear activations, using a combination of arithmetic and garbled circuits for a single semi-honest adversary in both the 2 and 3-server models. It also provided secure protocols for logistic sigmoid and softmax in the 2-server setting (although very briefly) and a method for truncating decimal numbers. MiniONN~\cite{liu2017oblivious} optimized the protocols of SecureML by reducing the offline cost of matrix multiplications and increasing the online cost for the 2-server model in the semi-honest adversary setting. Chameleon~\cite{riazi2018chameleon} and Gazelle~\cite{juvekar2018gazelle} provided secure inference protocols which are computationally secure against one semi-honest adversary in the 3-server and 2-server models. Chameleon removes expensive oblivious transfer protocols by using the third party as a dealer, while Gazelle focuses on making the linear layers more communication efficient by providing specialized packing schemes for additively homomorphic encryption schemes. 

SecureNN~\cite{wagh2018securenn} provides secure neural network training protocols for non-linear activation functions in the 3-party setting with up to one malicious corruption and shows that the honest-majority can improve the performance by several orders of magnitude. For linear regression, logistic regression and neural networks, the problem is even more challenging as the training procedure computes many instances of non-linear activation functions such as logistic sigmoid, tanh, and softmax that are expensive to compute inside a 2 and 3-server.

SecureNN provides efficient protocols for computing the ReLU activation, maxpool and their derivatives. It suggests that the general idea can extended to other variants and piecewise linear approximations. However, ~\cite{mohassel2017secureml} show experimentally that low-degree polynomial approximations, like piecewise linear approximations, are ineffective in terms of accuracy. It can be proven that hard sigmoid, a piecewise linear approximation of the logistic sigmoid, performs worse than logistic sigmoid, especially in the case of linear regression \cite{hardsig}. SecureML also provides higher-order approximations of logistic sigmoid and softmax  using a combination of ReLU functions in place of exponentiation. However, they do not provide a detailed algorithm and find the running time in the offline phase to be high. 

To overcome this research gap, we propose \textit{S++}, based on ~\cite{wagh2018securenn}, for the 3-server setting with efficient protocols for \textit{logistic sigmoid} and \textit{tanh} and their derivatives, as well as the \textit{softmax}. This can be extended to the derivative of the softmax as well.




%% file: background.tex
\section{Cryptographic Primitives}\label{section:background}

\subsection{Notation}\label{subsection:notation}
In our work, we use 2-out-of-2 additive sharing over the even ring $Z_{L}$ where $L = 2^l$. For our purposes, we use the same ring as~\cite{wagh2018securenn}, i.e., $Z_{2^{64}}$. 
The 2-out-of-2 additive shares are represented by $\langle x \rangle_0^L$ and $\langle x \rangle_1^L$. where L represents the ring $Z_L$ over which the value $x$ has been shared. 

We use the notation $\langle a \rangle_j$ to denote Party $P_j$'s additive secret share of the value a, and $\lfloor x \rfloor$ to denote the floor of the value x. The notation $(p_0, p_1, \cdots, p_{n-1})$ is used to denote the bits of a n-bit value, $p$.

Further, we use the following primitives from~\cite{wagh2018securenn} in our work: 

\begin{enumerate}
    \item Matrix Multiplication of two secret shared matrices represented by $\mathcal{F}_{MatMul}(\{P_0, P_1\} P_2)$.
    
    
    \item Division of a two secret shared values represented by $\mathcal{F}_{Division}(\{P_0, P_1\} P_2)$.
    
\end{enumerate}
These protocols are described in the following section.

%% file: protocol.tex
\section{Protocols of S++}\label{section:protocol}

\subsection{Architecture}\label{subsection:architecture}
S++ works with the $3$-server setting similar to SecureNN~\cite{wagh2018securenn}. In this setting, there are two primary servers and one helper server. The two primary servers hold additive secret shares of the data. Let $P_0, P_1, P_2$ be the three servers and $C$ be a set of $n$ participants $\{C_1, C_2, \dots C_n\}$, where the $i^{th}$ party $C_i$ holds its own private dataset, $\mathbb{D}_i$. $\mathcal{M}_{nn}$ is the neural network model to be trained on the parties' private data. The participants $\{C_1, C_2, \dots C_n\}$ split their data into 2-out-of-2 additive secret shares and give one share to each of the two servers $P_0$ and $P_1$. The third server, $P_2$, similar to ~\cite{wagh2018securenn}, is crucial to the protocols, but does not hold any data. We use the same integer representation for fixed point numbers as ~\cite{wagh2018securenn}. However, to tackle any overflow that can render the exponentiation protocol impractical for use, we also consider extending the integer ring to $Z_{128}$. S++ provides security against one semi-honest adversary and up to one malicious corruption.



\begin{figure}[ht]
    \centering
    \includegraphics[width=0.45\textwidth]{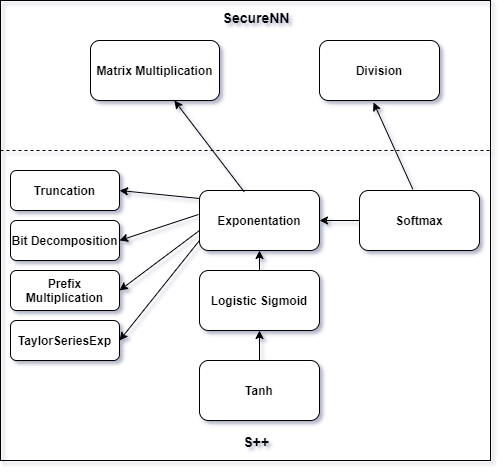}
    \caption{Function Dependencies in S++.}
    \label{fig:fundep}
\end{figure}




\subsection{Supporting Protocols}\label{subsection:supportprotocols}
\subsubsection{Matrix Multiplication ($\mathcal{F}_{MatMul}(\{P_0, P_1\} P_2)$):}

In S++, this function from~\cite{wagh2018securenn} has been used for multiplying fixed point values (represented as unsigned integers in the $Z_L$ ring) by considering the values as $1\times1$ matrices.

Two parties, $P_0$ and $P_1$ are required to hold shares of $X \in Z_L^{m\times n}$ and $Y \in Z_L^{n\times v}$. After running this interactive secure protocol, $P_0$ gets $\langle X.Y \rangle^L_0$ and $P_1$ gets $\langle X.Y \rangle^L_1$.

\subsubsection{Division ($\mathcal{F}_{Division}(\{P_0, P_1\} P_2)$):}

This function, also from ~\cite{wagh2018securenn}, is used to perform secure division on secret shared values. In this protocol, two parties, $P_0$ and $P_1$ are required to hold shares of values $x_j$ and $y_j$ respectively, for $j \text{in} \{0,1\}$. After running the protocol, they obtain $\langle x/y \rangle^L_0$ and $\langle x/y \rangle^L_1$ respectively.

\subsubsection{Taylor Series Expansion ($\mathcal{F}_{taylorExp}(\{P_0, P_1\} P_2)$; see algorithm~\ref{alg:tayloralg}):}

In this protocol, we compute $e^x$, where $x$ is a secret fractional value. We do so by computing the value of the Taylor expansion up to four terms. In the end of the protocol, parties $P_0$ and $P_1$ obtain shares of $e^x$.
\begin{algorithm}
\caption{Taylor Expansion, $\mathcal{F}_{taylorExp}(\{P_0, P_1\},P_2)$}
\label{alg:tayloralg}
\hspace*{\algorithmicindent} \textbf{Input:} $P_0, P_1$ hold $\{\langle x \rangle_0^L\}$ and $\{\langle x \rangle_1^L\}$ respectively where $x<1$.

\hspace*{\algorithmicindent} \textbf{Output:} $P_0$ and $P_1$ obtain  $\langle y \rangle_j^L$ = $\langle e^{x} \rangle _j^L$.

\hspace*{\algorithmicindent} \textbf{Common Randomness:} $P_0$ and $P_1$ hold random shares of zero - $u_0$ and $u_1$ respectively.  

\begin{algorithmic}[1]
      \STATE Parties $P_0$ and $P_1$ compute $\langle c \rangle_j^L = j$.
      
      \STATE Parties $P_0$ and $P_1$ compute $\langle numerator \rangle_j^L = \langle x \rangle_j^L$ and $denominator = 1$.
      
      \STATE Now, Parties $P_0$ and $P_1$ compute $\langle c \rangle_j^L = \langle c \rangle_j^L + \langle numerator \rangle_j^L$.
      
      \FOR{$i=2,\ldots,4$}
                \STATE Parties $P_0$, $P_1$ and $P_2$ invoke $\mathcal{F}_{MatMul}(\{P_0, P_1\} P_2)$ with inputs $\langle numerator \rangle_j^L$ and $\langle x \rangle_j^L$ to obtain $\langle numerator \rangle_j^L$ for $j \in \{0,1\}$.
                \STATE $denominator = denomiator \times i$
                \STATE Parties $P_0$ and $P_1$ compute $\langle c \rangle_j^L = \langle c \rangle_j^L + \frac{\langle numerator \rangle_j^L}{denominator}$
        \ENDFOR
      
      \STATE $P_j $ for $j \in \{0,1\}$ output $\langle c \rangle_j^L$ + $u_j$.
\end{algorithmic}
\end{algorithm}

\subsubsection{Exponentiation ($\mathcal{F}_{Exp}(\{P_0, P_1\},P_2)$); see algorithm~\ref{alg:exponentiation}:}

In this protocol, we compute $e^x$, where $x$ is secret. This protocol is based on SCALE MAMBA's~\cite{aly2020scale} base-2 exponentiation protocol. It uses primitives like $\mathcal{F}_{Trunc}$, $\mathcal{F}_{BitDecomp}$, $\mathcal{F}_{PreMult}$ from \cite{catrinasaxena}.

While \cite{aly2020scale} uses the polynomial $P_{1045}(X)$ from \cite{hart} to compute the exponentiation for the fractional part of a number, we use a secure protocol to compute the output using the Taylor series expansion of $e^x$. We introduce function $\mathcal{F}_{taylorExp}$ defined above for this purpose.

This protocol first uses $\mathcal{F}_{Trunc}$ to split the input $a$ into its integer ($a_\text{int}$) and fractional ($a_\text{frac}$) parts. After that, it evaluates $e^{a_\text{int}}$ using primitives $\mathcal{F}_{BitDecomp}$ and $\mathcal{F}_{PreMult}$ described above. It further evaluates $e^{a_\text{frac}}$ using $\mathcal{F}_{taylorExp}$. It then multiplies these two values to obtain $e^{a_\text{int}+a_\text{frac}} = e^a$.

\begin{algorithm}
\caption{Exponentiation $\mathcal{F}_{Exp}(\{P_0, P_1\},P_2)$}
\label{alg:exponentiation}
\hspace*{\algorithmicindent} \textbf{Input:} $P_0$ and $P_1$ hold $\langle x \rangle_0^L$ and $\langle x \rangle_1^L$ (shares of a value x).

\hspace*{\algorithmicindent} \textbf{Output:} $P_0$ and $P_1$ obtain  $\langle y \rangle_j^L$ = $\langle e^{x} \rangle _j^L$.


\begin{algorithmic}[1]
      \STATE For $j \in \{0,1\}$, party $P_j$ calls $\mathcal{F}_{Trunc}(\{P_0, P_1\})$ to obtain $\langle a \rangle_j^L$, which is the $j^{\text{th}}$ share of $\lfloor x \rfloor$.

      \STATE For $j \in \{0,1\}$, party $P_j$ gets the fractional part of x by locally computing:
      \begin{center}
          $\langle b \rangle_j^L = \langle x \rangle_j^L - \langle a \rangle_j^L$
      \end{center}

      \STATE For $j \in \{0,1\}$ party $P_j$ invokes $\mathcal{F}_{BitDecomp}(\{P_0, P_1\})$ to obtain $(\langle c_0\rangle_j^L, \langle c_1\rangle_j^L, \cdots, \langle c_{m-1}\rangle_j^L)$: shares of $(x_0, x_1, \cdots, x_{m-1})$, where $x$ is a m-bit number. 
      
    \FOR{$i=0,1,\ldots,m-1$}
            \STATE $P_j$ for $j \in \{0,1\}$ computes $\langle v_i \rangle_j^L = e^{2^i}.(\langle c_i\rangle_j^L) + j - (\langle c_i\rangle_j^L)$.
    \ENDFOR
      \STATE $P_j$ for $j \in \{0,1\}$ invokes $\mathcal{F}_{PreMult}(\{P_0, P_1\})$ to get $\langle m \rangle_j^L$.
      
      \STATE $P_j$ for $j \in \{0,1\}$ invokes $\mathcal{F}_{Taylor}(\{P_0, P_1\})$ to get $\langle n\rangle_j^L$
      
      \STATE Finally, $P_j$ for $j \in \{0,1\}$ invokes $\mathcal{F}_{MatMul}(\{P_0, P_1\} P_2)$ with inputs $\langle m\rangle_j^L$ and $\langle n\rangle_j^L$ to obtain share $\langle m\rangle_j^L$ of:
      \begin{center}
          $y = m\times n$.
      \end{center}
\end{algorithmic}
\end{algorithm}

\subsection{Main Protocols}\label{subsection:mainprotocols}
Our main protocols include logistic sigmoid ($\mathcal{F}_{Sigmoid}$; described in algorithm~\ref{alg:sigmoid}), tanh ($\mathcal{F}_{Tanh}$; described in algorithm~\ref{alg:tanh}), and their derivatives; see algorithms~\ref{alg:derivsigmoid} and ~\ref{alg:derivtanh}, as well as the softmax function ($\mathcal{F}_{Softmax}$; described in algorithm~\ref{alg:softmax}).

\subsubsection{Sigmoid, $\mathcal{F}_{Sigmoid}(\{P_0, P_1\},P_2)$; see algorithm~\ref{alg:sigmoid}):}
This protocol can be used to securely compute logistic sigmoid. Parties $P_0$, $P_1$ and $P_2$ invoke $\mathcal{F}_{Exp}$ with shares of the input $x$ to obtain output shares of $e^x$. They then jointly compute shares of $1+e^x$. They use~\cite{wagh2018securenn}'s secure division protocol to obtain $\sigma(x) = \frac{e^x}{1+e^x}$.

\begin{algorithm}[H]
\caption{Sigmoid, $\mathcal{F}_{Sigmoid}(\{P_0, P_1\},P_2)$}
\label{alg:sigmoid}
\hspace*{\algorithmicindent} \textbf{Input:} $P_0, P_1$ hold $\{\langle x \rangle_0^L\}$ and $\{\langle x \rangle_1^L\}$ respectively.

\hspace*{\algorithmicindent} \textbf{Output:} $P_0, P_1$ get $\{\langle \sigma(x) \rangle_0^L\}$ and $\{\langle \sigma(x) \rangle_1^L\}$ respectively where $\sigma(x) = \frac{e^x}{1 + e^x}$.

\hspace*{\algorithmicindent} \textbf{Common Randomness:} $P_0$ and $P_1$ hold random shares of zero - $u_0$ and $u_1$ respectively.  

\begin{algorithmic}[1]
      \STATE Parties $P_0$, $P_1$ and $P_2$ invoke $F_{Exp}(\{P_0,P_1\},P_2)$  with inputs $\langle x \rangle_0^L$ and $\langle x \rangle_1^L$ and obtain $\langle a \rangle_j^L = \langle e^x \rangle_j^L$.
      
      \STATE Now, parties $P_0$ and $P_1$ compute $\langle b \rangle_j^L = \langle a \rangle_j^L + j$.
      
      \STATE $P_0$, $P_1$ and $P_2$ invoke $\mathcal{F}_{Division}(\{P_0,P_1\},P_2)$ with inputs $\langle a \rangle_j^L$ and $\langle b \rangle_j^L$ to obtain $\langle c \rangle_j^L$ for $j \in \{0,1\}$.
      
      \STATE $P_j $ for $j \in \{0,1\}$ output $\langle c \rangle_j^L$ + $u_j$.
\end{algorithmic}
\end{algorithm}
\subsubsection{Derivative of Sigmoid, $\mathcal{F^D}_{Sigmoid}(\{P_0, P_1\},P_2)$; see algorithm~\ref{alg:derivsigmoid}):}
This protocol can be used to securely compute the derivative of the logistic sigmoid. We use the idea that $\frac{dS (x)}{d(x)} = \sigma'(x) = \sigma(x)\cdot (1-\sigma(x)).$ In the end of the protocol, parties $P_0$ and $P_1$ obtain the shares of $\sigma'(x)$.
\begin{algorithm}[H]
\caption{Derivative of Sigmoid $\langle{\mathcal{F^D}_{sigmoid}(x)}\rangle^L$}
\label{alg:derivsigmoid}
\hspace*{\algorithmicindent} \textbf{Input:} $P_0$, $P_1$ have inputs $\langle{x}\rangle_j^L$ for $P_j$ such that $j \in \{0,1\}$.

\hspace*{\algorithmicindent} \textbf{Output:} $P_0$, $P_1$ get $\langle{\sigma'(c)}\rangle_0^L$ and $\langle{\sigma'(c)}\rangle_1^L$ where $\sigma'(c) = \sigma(x)(1-\sigma(x))$

\hspace*{\algorithmicindent} \textbf{Common Randomness:} $P_0$ and $P_1$ hold random shares of zero - $u_0$ and $u_1$ respectively. 

\begin{algorithmic}[1]
      \STATE $P_0$, $P_1$ and $P_2$ invoke $\mathcal{F}_{Sigmoid}{(\{P_0, P_1\}, P_2)}$ with $P_j$ for $j \in \{0,1\}$ having input $\langle{x}\rangle_j^L$ to learn $\langle{a}\rangle_j^L =  \langle{\sigma(x)}\rangle_j^L$.
      
      \STATE $P_j$ computes $\langle{a}\rangle_j^L$ = $-1 \times \langle{x}\rangle_j^L + j$ for $j \in \{0,1\}$
      
      \STATE $P_j$ computes $\langle{b}\rangle_j^L$ = $\langle{a}\rangle_j^L$ for $j \in \{0,1\}$
      
      \STATE $P_0$, $P_1$ and $P_2$ invoke $\mathcal{F}_{MatMul}{(\{P_0, P_1\}, P_2)}$ with $P_j$ for $j \in \{0,1\}$ having inputs $\langle{a}\rangle_j^L$ and $\langle{b}\rangle_j^L$ to learn $\langle{c}\rangle_j^L =  \langle{a}\rangle_j^L \times \langle{b}\rangle_j^L$.
      
      \STATE $P_j$ for $j \in \{0,1\}$ output:
            \begin{center}
                $\langle{c}\rangle_j^L + \langle{u}\rangle_j^L$
            \end{center}
      
\end{algorithmic}
\end{algorithm}

\subsubsection{Tanh, $\mathcal{F}_{Tanh}(\{P_0, P_1\},P_2)$; see algorithm~\ref{alg:tanh}):}
This protocol can be used to securely compute the tanh function. For this protocol, we use the idea that $tanh(x)$ is related to $\sigma(x)$ as $tanh(x) = 2 \sigma(2x) -1$.

\begin{algorithm}[h]
\caption{Tanh $\langle{\mathcal{F}_{Tanh}(x)}\rangle^L$}
\label{alg:tanh}

\hspace*{\algorithmicindent} \textbf{Input:} $P_0$, $P_1$ have inputs $\langle{x}\rangle_j^L$ for $P_j$ such that $j \in \{0,1\}$.

\hspace*{\algorithmicindent} \textbf{Output:} $P_0$, $P_1$ get $\langle{tanh(c)}\rangle_0^L$ and $\langle{tanh(c)}\rangle_1^L$ where $tanh(c) = \frac{2}{1 + e^{-2x}} - 1$

\hspace*{\algorithmicindent} \textbf{Common Randomness:} $P_0$ and $P_1$ hold random shares of zero - $u_0$ and $u_1$ respectively. 

\begin{algorithmic}[1]
      \STATE $P_j$ computes $\langle{a}\rangle_j^L$ = $2 \times \langle{x}\rangle_j^L$ for $j \in \{0,1\}$

      \STATE $P_0$, $P_1 and P_2$ invoke $\mathcal{F}_{Sigmoid}{(\{P_0, P_1\}, P_2)}$ with $P_j$ for $j \in \{0,1\}$ having input $\langle{a}\rangle_j^L$ to learn $\langle{p}\rangle_j^L =  \langle{\sigma(a)}\rangle_j^L$.
      
      \STATE $P_j$ for $j \in \{0,1\}$ output:
            \begin{center}
                $2 \times \langle{P}\rangle_j^L - j + \langle{u}\rangle_j^L$
            \end{center}
      
\end{algorithmic}
\end{algorithm}

\subsubsection{Derivative of Tanh, $\mathcal{F^D}_{Tanh}(\{P_0, P_1\},P_2)$; see algorithm~\ref{alg:derivtanh}):}
This protocol can be used to securely compute the derivative of the tanh function. For this protocol, we use the idea that $tanh'(x)$ is related to $\sigma'(x)$ as $tanh'(x) = 4\sigma'(2x)$.
\begin{algorithm}
\caption{Derivative of Tanh $\langle{\mathcal{F^D}_{Tanh}(x)}\rangle^L$}
\label{alg:derivtanh}
\hspace*{\algorithmicindent} \textbf{Input:} $P_0$, $P_1$ have inputs $\langle{x}\rangle_j^L$ for $P_j$ such that $j \in \{0,1\}$.

\hspace*{\algorithmicindent} \textbf{Output:} $P_0$ and $P_1$ get $\langle{\tanh'(c)}\rangle_0^L$ and $\langle{\tanh'(c)}\rangle_1^L$ respectively.

\hspace*{\algorithmicindent} \textbf{Common Randomness:} $P_0$ and $P_1$ hold random shares of zero - $u_0$ and $u_1$ respectively. 

\begin{algorithmic}[1]
      
      
    \STATE $P_0$ and $P_1$ compute $\langle{a}\rangle_j^L = 2\times \langle{x}\rangle_j^L$.
    \STATE $P_0$, $P_1$, $P_2$ invoke $\mathcal{F^D}_{sigmoid}$ with $\langle{a}\rangle_j^L$ to obtain shares $\langle{b}\rangle_j^L$.
    \STATE $P_0$ and $P_1$ output:
    \begin{center}
        $4.\langle{b}\rangle_j^L + \langle{u}\rangle_j^L$
    \end{center}
\end{algorithmic}
\end{algorithm}

\subsubsection{Softmax, $\mathcal{F}_{Softmax}(\{P_0, P_1\},P_2)$; see algorithm~\ref{alg:derivtanh}):}
This protocol can be used to securely compute the softmax function. The two parties compute the numerator $e^{z_i}$ and the denominator $\sum_{i=1}^k e^{z_i}$ and invoke the $\mathcal{F}_{Division}$ function to perform secure division.
\begin{algorithm}
\caption{Softmax, $\mathcal{F}_{Softmax}(\{P_0, P_1\},P_2)$}
\label{alg:softmax}
\hspace*{\algorithmicindent} \textbf{Input:} $P_0, P_1$ hold $\{\langle z_i \rangle_0^L\}_{i\in[k]}$ and $\{\langle z_i \rangle_1^L\}_{i\in[k]}$ respectively.

\hspace*{\algorithmicindent} \textbf{Output:} $P_0, P_1$ get $\{\langle s_{max}(z_i) \rangle_0^L\}_{i\in[k]}$ and $\{\langle s_{max}(z_i) \rangle_1^L\}_{i\in[k]}$ respectively where $s_{max}(z_i) = \frac{e^{z_i}}{\sum_{i=1}^k e^{z_i}}$.

\hspace*{\algorithmicindent} \textbf{Common Randomness:} $P_0$ and $P_1$ hold random shares of zero - $u_0$ and $u_1$ respectively.

\begin{algorithmic}[1]
\FOR{$i=1,2,\ldots,k$}
      \STATE Parties $P_0$, $P_1$ and $P_2$ invoke $F_{Exp}(\{P_0,P_1\},P_2)$  with inputs $\{\langle z_i \rangle_0^L\}_{i\in[k]}$ and $\{\langle z_i \rangle_1^L\}_{i\in[k]}$ and $P_0$, $P_1$ obtain shares $\langle c_i \rangle_0^L$, $\langle c_i \rangle_1^L$ resp. of $c_i^L$ = $e^{z_i}$.
\ENDFOR

\STATE for $j \in \{0,1\}$, $P_j$ calculates $S_j = \sum_{i=1}^k \langle c_i \rangle_j^L$.


\FOR{$i=1,2,\dots,k$}
      \STATE Parties $P_0$, $P_1$ and $P_2$ invoke $\mathcal{F}_{Division}(\{P_0,P_1\},P_2)$ with inputs $\langle c_i \rangle_j^L$ and $\langle S \rangle_j^L$.

      \STATE Parties $P_j$ for $j \in \{0,1\}$ output $\{\langle S_{max
      }(z_i) \rangle_j^L\}_{i \in [k]} + u_j$.
\ENDFOR
\end{algorithmic}
\end{algorithm}










%% file: evaluation.tex
\section{Empirical Evaluation}\label{section:eval}


\subsection{Implementation and Experimental Setup}\label{subsection:implement}
Our system is implemented in C++ using standard libraries. To be able to execute machine learning protocols, we use fixed-point representation in the integer ring $\mathbb{Z}_{64}$. This allows us to compute precise values using our machine learning protocols. The fixed-arithmetic used in our implementation is the same as \cite{wagh2018securenn} which borrowed the idea from \cite{mohassel2017secureml}.

Exponentiation is complicated and may overflow the integer ring. So, we consider the ring $\mathbb{Z}_{128}$ as proposed by \cite{aly2020scale}. So far, we have implemented the Taylor series exponentiation (refer to step 7 of Algorithm~\ref{alg:tayloralg}) which works for values less than 1.  We provide benchmarks for our logistic sigmoid, tanh and softmax functions, their derivatives and the Taylor series exponentiation in the secure setting with input vectors of varying sizes. These benchmarks include the execution time and the total communication (in MB) of these protocols.

We run our protocols in a LAN setting on an Intel i5 7th Gen with 8GB RAM. The average bandwidth measured was 40Mb/s for download and 9.2Mb/s for upload and the average ping was 12ms.

\subsection{Experimental Results}\label{subsection:result}
We show the average time taken by the protocols in Table~\ref{tab:benchmarks}.

\begin{table}[H]
    \centering
    \begin{tabular}{c c c c}
        \hline
         Protocol & Dimension & Time(s) & Comm.(mb) \\
        \hline
                                    & 64x16 & 0.08 & 0.025 \\
         $\mathcal{F}_{Exp}$        & 128x128 & 2.134 & 0.393 \\
                                    & 576x20 & 0.882 & 0.276 \\
        \hline
                                    & 64x16 & 0.252 & 2.58 \\
         $\mathcal{F}_{Sigmoid}$    & 128x128 & 5.631 & 41.288 \\
                                    & 576x20 & 2.615 & 29.03 \\
        \hline
                                    & 64x16 & 0.275 & 2.58 \\
         $\mathcal{F}_{tanh}$       & 128x128 & 5.32 & 41.288 \\
                                    & 576x20 & 2.613 & 29.03 \\
        \hline
                                    & 64x16 & 0.324 & 2.58 \\
         $\mathcal{F}_{Softmax}$    & 128x128 & 5.438 & 41.288 \\
                                    & 576x20 & 2.617 & 29.03 \\
        \hline
                                    & 64x16 & 0.464 & 2.597 \\
         $\mathcal{F^D}_{sigmoid}$  & 128x128 & 8.033 & 41.55 \\
                                    & 576x20 & 4.121 & 29.214 \\
        \hline
                                    & 64x16 & 0.383 & 2.58 \\
         $\mathcal{F^D}_{tanh}$     & 128x128 & 4.465 & 41.288 \\
                                    & 576x20 & 2.84 & 29.03 \\
        \hline
                                    & 64x16 & 0.032 & 0.005 \\
         $\mathcal{F}_{taylorExp}$  & 128x128 & 0.092 & 0.079 \\
                                    & 576x20 & 0.427 & 0.055 \\
        \hline
    \end{tabular}
    \caption{Benchmarks in LAN setting on i5 7th Gen}
    \label{tab:benchmarks}
\end{table}


    
    
    



%% file: conclusion.tex
\section{Conclusion and Future Work}\label{section:conclusion}

Previous work in secure computation of neural networks has been able to achieve reasonable efficiency for many real world applications. However, the lack of options in the choice of protocols that these implementations can run can limit the capacity of activities the user is able to perform in the secure setting.

Through the addition of the exponentiation function, it becomes possible to add a large number of secure activation functions that were previously impossible to compute. We have shown logistic sigmoid, tanh, softmax and their derivatives as examples of such functions. We implement these functions with similar overheads as the previously proposed activation functions.

In its current form, the exponentiation function overflows for small values making it impractical for usage in regular neural network. This also makes us unable to use the functions that call the exponentiation function, as part of their execution, in practical settings.

Exponentiation protocols in a fixed-point MPC setting have been provided by \cite{aly2019benchmarking} and implemented by \cite{aly2019zaphod}. Ideas from these protocols can be used to avoid using a workaround that would involve increasing the size of shares to accommodate the overflow.

The future work on S++ will be focused on dealing with the overflow such that the exponentiation function can handle the magnitude of the values that it would encounter in a regular neural network. Once this is solved, the other functions that we have proposed, that use exponentiation, will automatically become suitable for practical usage.

%% file: appendix.tex
\section{Appendix}\label{section:appendix}

\textbf{Universal approximation of softmax: }\cite{asadi2020approximation} provide theoretical proofs on the universal approximation of ReLU used in hidden layers of deep networks and softmax used in the output layer for multiclass classification tasks. They find that such a proof for the softmax is missing in the literature despite it being widely used for these tasks, and provide their own to prove the power of the softmax used in output layers to approximate indicator functions.\\

\textbf{ReLU and gradients: }\cite{pascanu2013difficulty} delve into the exploding and vanishing gradients of activation functions that tend to cause problems in recurrent neural networks because of the temporal correlations they capture (by forming connections between hidden layers). Long-term components add up and the norm of the gradient of the hidden layers \textit{explode}. ReLU's unbounded nature exacerbates this problem. There have been a few notable improvements to assuage this, namely gradient clipping, identity initialization \cite{le2015simple}, and gated RNNs like LSTMs and GRUs. However, even in LSTMs, the tanh often gives more consistent results (\cite{kent2019performance}) and ReLU tends to require greater fine-tuning. \\

\textbf{The Leaky Tanh: }\cite{xu2016revise} propose a `leaky' tanh that penalizes the negative part i.e., instead of the regular tanh function, the penalized tanh behaves like this:
\begin{equation}
    penalized\_tanh(x) = \begin{dcases}
                        tanh(x) & x > 0 \\
                        a.tanh(x) & otherwise \\
                        \end{dcases}
\end{equation}
where $a \in (0, 1)$. They also found this variation to be two times faster than the tanh.
